\documentstyle[amssymb,prb,aps,epsfig,multicol]{revtex}  
\begin{document}
\title{Electronic Structure and Electron-Phonon Coupling in
the 18K Superconductor Y$_2$C$_3$}
\author{D.J. Singh and I.I. Mazin}
\address{Center for Computational Materials Science,
Naval Research Laboratory, Washington, DC 20375}
\date{\today}
\maketitle

\begin{abstract}
The electronic structure and electron-phonon coupling in Y$_2$C$_3$
is investigated using density functional calculations. We find that
the Fermi level falls in a manifold of mixed character derived from
Y $d$ states and antibonding states associated with the C dimers
in the structure. Calculations of the electron-phonon coupling
for Y and C modes show that the former provide most of the coupling.
Modes associated with C-C bond stretching have large matrix elements,
but make small
contributions to the coupling because of their
high phonon frequencies. Substantial
electron doping of the C-C antibonding states
would yield a large increase in the coupling and critical temperature,
perhaps to values comparable to MgB$_2$. However, it seems unlikely
that a modification of Y$_2$C$_3$ with much higher
filling of the C-C antibonding states can be stabilized.
\end{abstract}

\pacs{74.25.Jb,74.70.Ad}

\begin{multicols}{2}


The rare earth sesquicarbides, $R_2$C$_3$, where $R$ is a rare earth or Y
are a family of superconducting
materials that form in the bcc Pu$_2$C$_3$ structure under high temperature
conditions.
Samples produced by arc melting have critical temperatures, $T_c$, of
10K to 12K, with little dependence on the particular rare earth.
For example, as synthesized samples with the largest and smallest
rare earths, specifically, La$_2$C$_3$ and Y$_2$C$_3$ have critical
temperatures of 11K and 10.5K, respectively.
\cite{giorgi0,giorgi,franc,ver-2,vereshchagin}
However, it was found early on that heavy Th doping raises $T_c$,
up to 17K in the case of the Y compound. \cite{krupka}
Very recently, Amano and co-workers \cite{amano} reported that $T_c$ for
Y$_2$C$_3$ could be increased to 18K by synthesis under high
pressure (4-5.5 GPa). The resulting samples have the same Pu$_2$C$_3$
structure as previously obtained samples but a slightly different lattice
parameter, suggesting that the effect of pressure is to change the
composition of the samples. Furthermore, they reported that
the results were sensitive to the sintering conditions. The lattice
parameters varied in the range $a$=8.181\AA ~-- 8.226\AA, with reported
$T_c$ spanning the range 15K to 18K. The extrapolated upper critical field
of the 18K samples was estimated to be $\sim$ 30T. \cite{nakane}

\begin{figure}[tbp]
\centerline{\epsfig{file=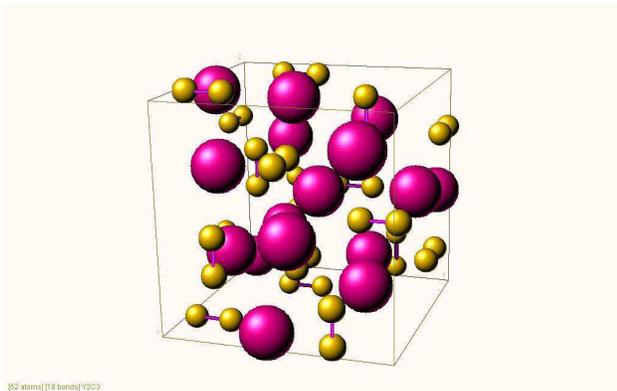,width=0.95\linewidth,angle=0,clip=}}
\vspace{0.2cm}
\caption{Crystal structure of Y$_2$C$_3$ showing the C$_2$ dimers.
Large spheres denote Y and small spheres denote C.}
\label{str}
\end{figure}

The Pu$_2$C$_3$ structure has four formula units per primitive unit
cell, spacegroup $I\bar{4}3d$, Y on site 16c ($u$,$u$,$u$) and
C on site 24d ($v$,0,1/4), where the coordinates are in terms of the
bcc lattice vectors (see Fig. \ref{str}). As may be seen, C
dimers with short bond lengths are a characteristic of this structure.
Electronic structure studies of other metals with C dimers both
at the tight binding \cite{lee} and density functional level \cite{gulden}
suggest that the electronic structure
could have substantial C-C antibonding character near the Fermi level.
Such states may be expected to have extremely high deformation potentials,
reflecting these very strong C-C triple bonds. Also reflecting the very
strong covalent bonds, phonons that modulate the C-C distances would
be very stiff,
which, if these states were responsible for the superconductivity,
would yield a moderate electron-phonon coupling, $\lambda$,
and a very high logarithmic average phonon frequency and therefore
high prefactor in the McMillan equation for $T_c$. In this scenario,
a very strong dependence of $T_c$ on stoichiometry (more precisely,
occupation of the antibonding states) might be anticipated,
as well as the possibility of very high $T_c$ in more optimally doped
samples, perhaps with values like in MgB$_2$ and $A_3$C$_{60}$.
Here we report density functional calculations of the electronic
structure and the coupling constants of a C-C and an Y mode
aimed at determining
the extent to which this scenario applies in Y$_2$C$_3$.

The present calculations were done using the general potential
linearized augmented planewave (LAPW) method, \cite{singh-book}
with local orbital extensions \cite{singh-lo}
to relax linearization errors and to treat the semi-core states of Y.
LAPW sphere radii of 2.0 $a_0$ and 1.24 $a_0$ were used for Y and C,
respectively. Well converged basis sets (convergence tests were done)
consisting of 4250 basis functions for the 20 atom primitive cell were
used, along with converged Brillouin zone samplings.
The core states were treated relativistically and valence states
semirelativistically.
For the crystal structure we used the reported bcc $I\bar{4}3d$ Pu$_2$C$_3$
structure with the experimentally determined lattice parameter,
$a=8.226$ \AA,
from Ref. \onlinecite{amano}
but relaxed the internal coordinates using the calculated
atomic forces.
We found significant changes in the C-C dimer bond length from
the available experimental refinement (Ref. \onlinecite{exp-struct}).
Specifically, we obtain a C-C distance of 1.33 \AA,
and a Y-C distance of 2.51 \AA,
corresponding to internal parameters, $u$=0.0504 and $v$=0.2940.
The corresponding full symmetry Raman modes are an Y dominated mode
at 175 cm$^{-1}$ and an almost pure C-C bond stretching mode at
1442 cm$^{-1}$.
It seems likely that the difference between our results
and the C position of Ref. \onlinecite{exp-struct} is related to difficulty
in the refinement due to sample
stoichiometry, which is an issue in these sesquicarbides, as mentioned
above.

\begin{figure}[tbp]
\centerline{\epsfig{file=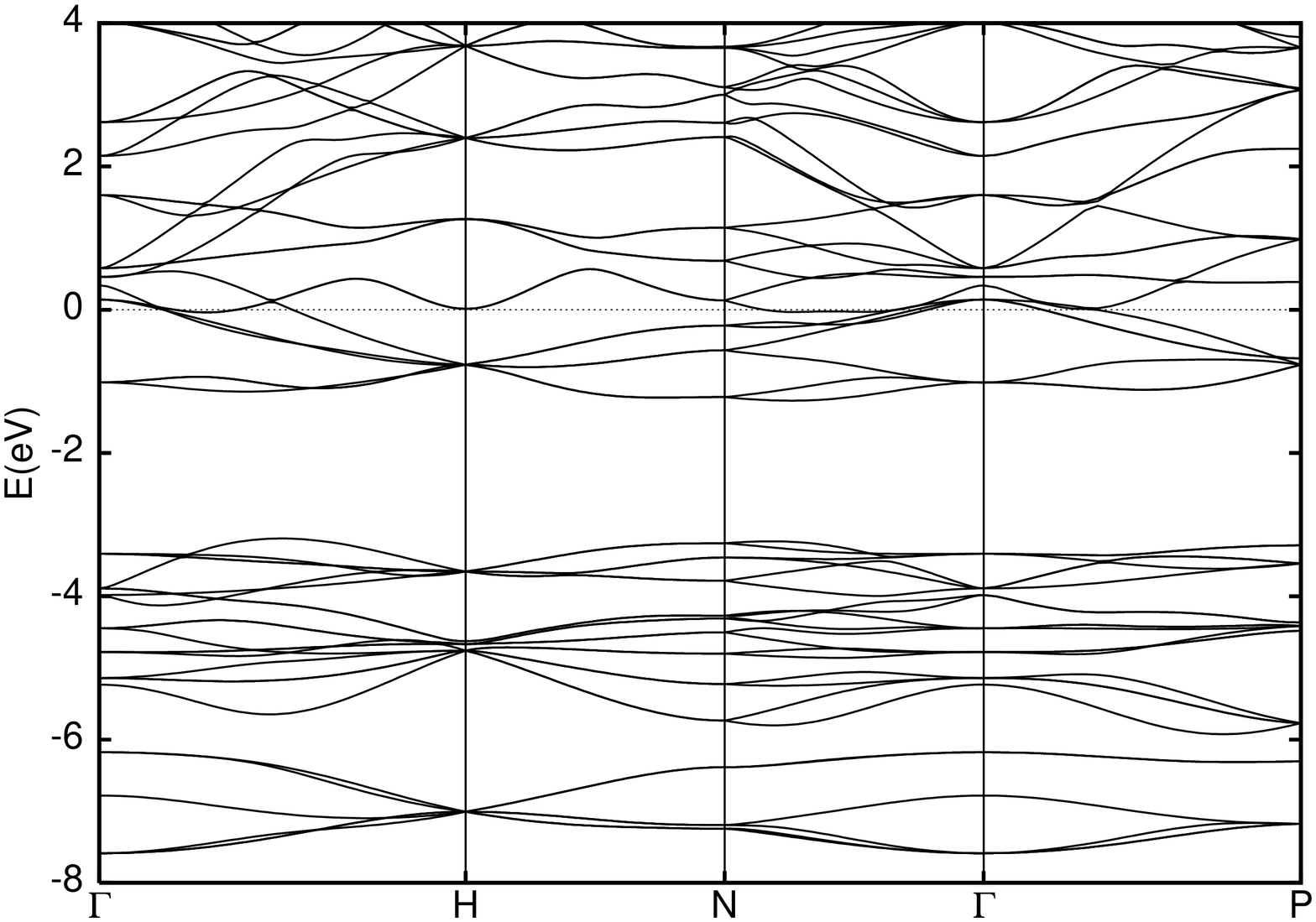,width=0.60\linewidth,angle=270,clip=}}
\vspace{0.3cm}
\centerline{\epsfig{file=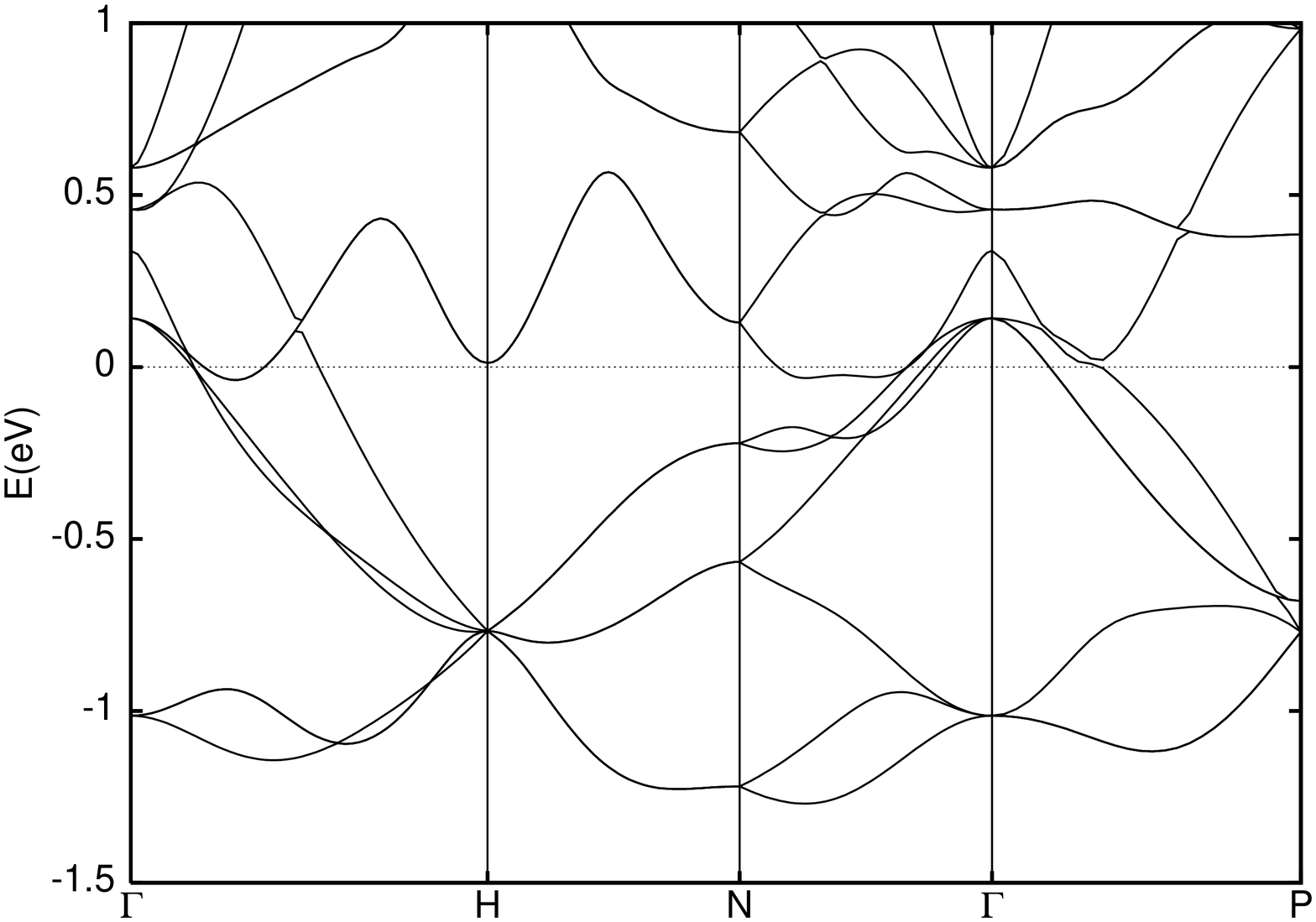,width=0.60\linewidth,angle=270,clip=}}
\vspace{0.3cm}
\caption{Calculated band structure of Y$_2$C$_3$ with the
relaxed atomic positions. The lowest bands shown are
part of the C 2$s$ manifold. The bottom of this
manifold is at -14.9 eV. The lower panel is a blow-up
around $E_F$, which is at 0 eV.}
\label{bands}
\end{figure}

\begin{figure}[tbp]
\centerline{\epsfig{file=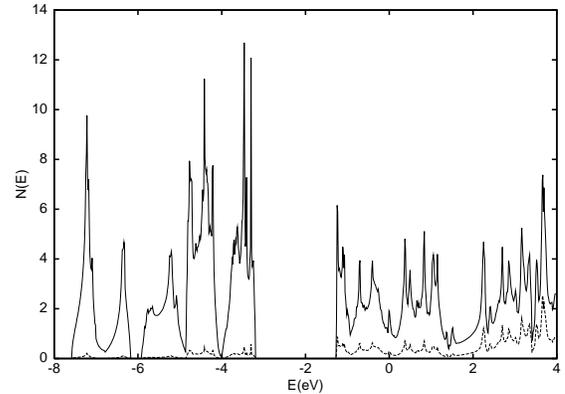,width=0.60\linewidth,angle=270,clip=}}
\vspace{0.3cm}
\caption{Electronic density of states of Y$_2$C$_3$ with the
relaxed atomic positions. The solid line is the total density of
states on a per formula unit basis and the dashed line is the
Y $d$ contribution as defined by projection onto the Y LAPW sphere
of radius 2.0 $a_0$.}
\label{dos}
\end{figure}

\begin{figure}[tbp]
\centerline{\epsfig{file=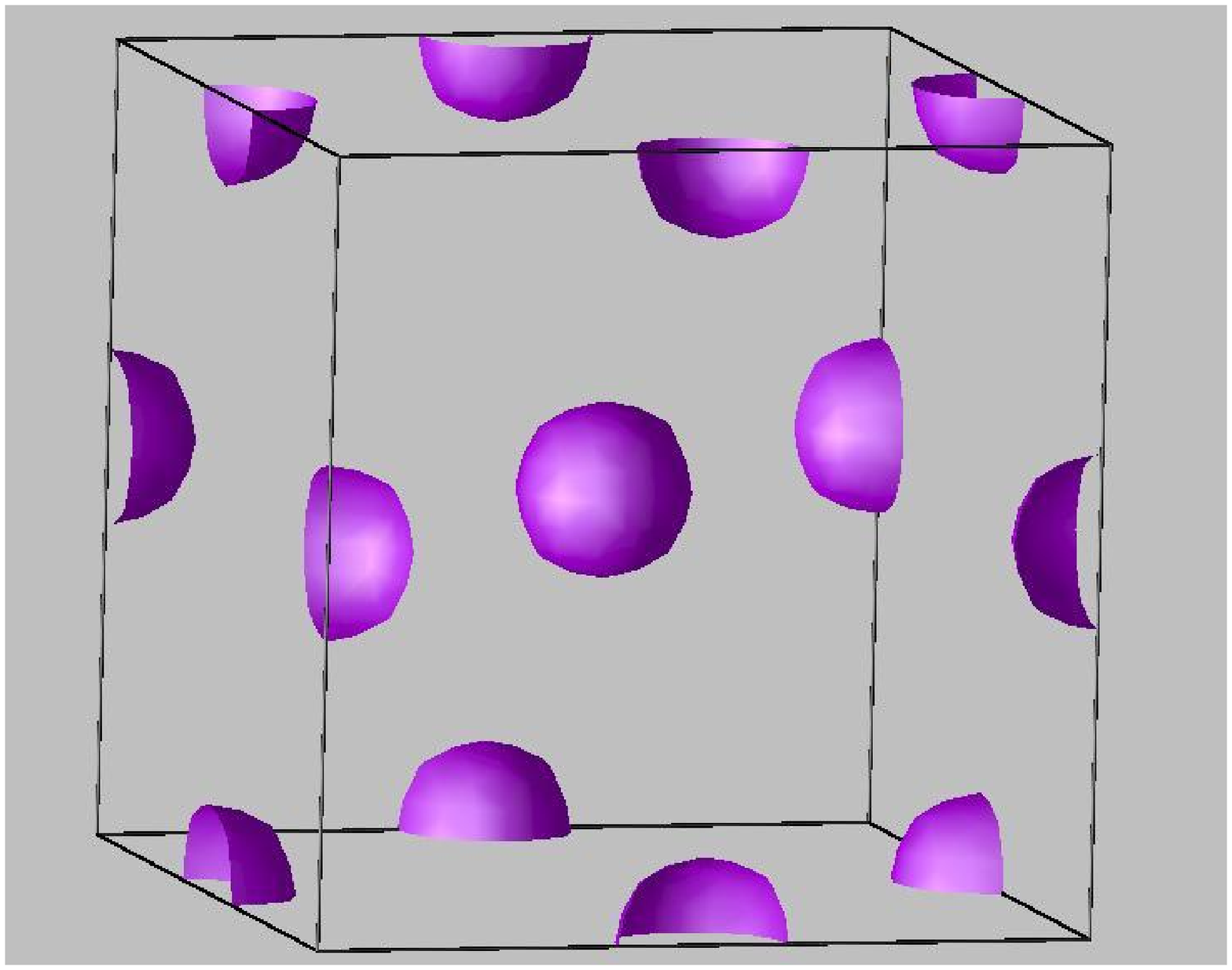,width=0.70\linewidth,angle=0,clip=}}
\vspace{0.2cm}
\centerline{\epsfig{file=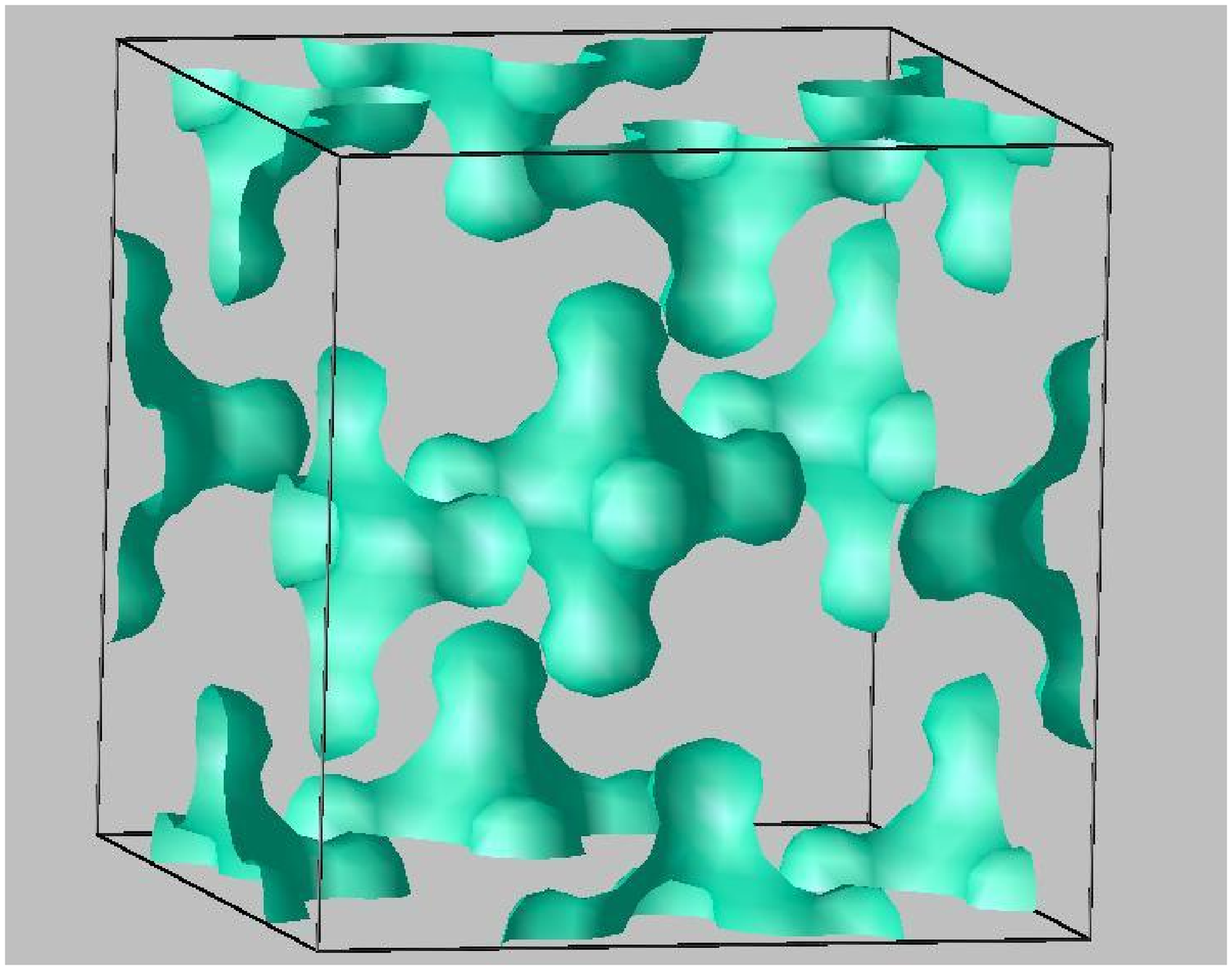,width=0.70\linewidth,angle=0,clip=}}
\vspace{0.2cm}
\centerline{\epsfig{file=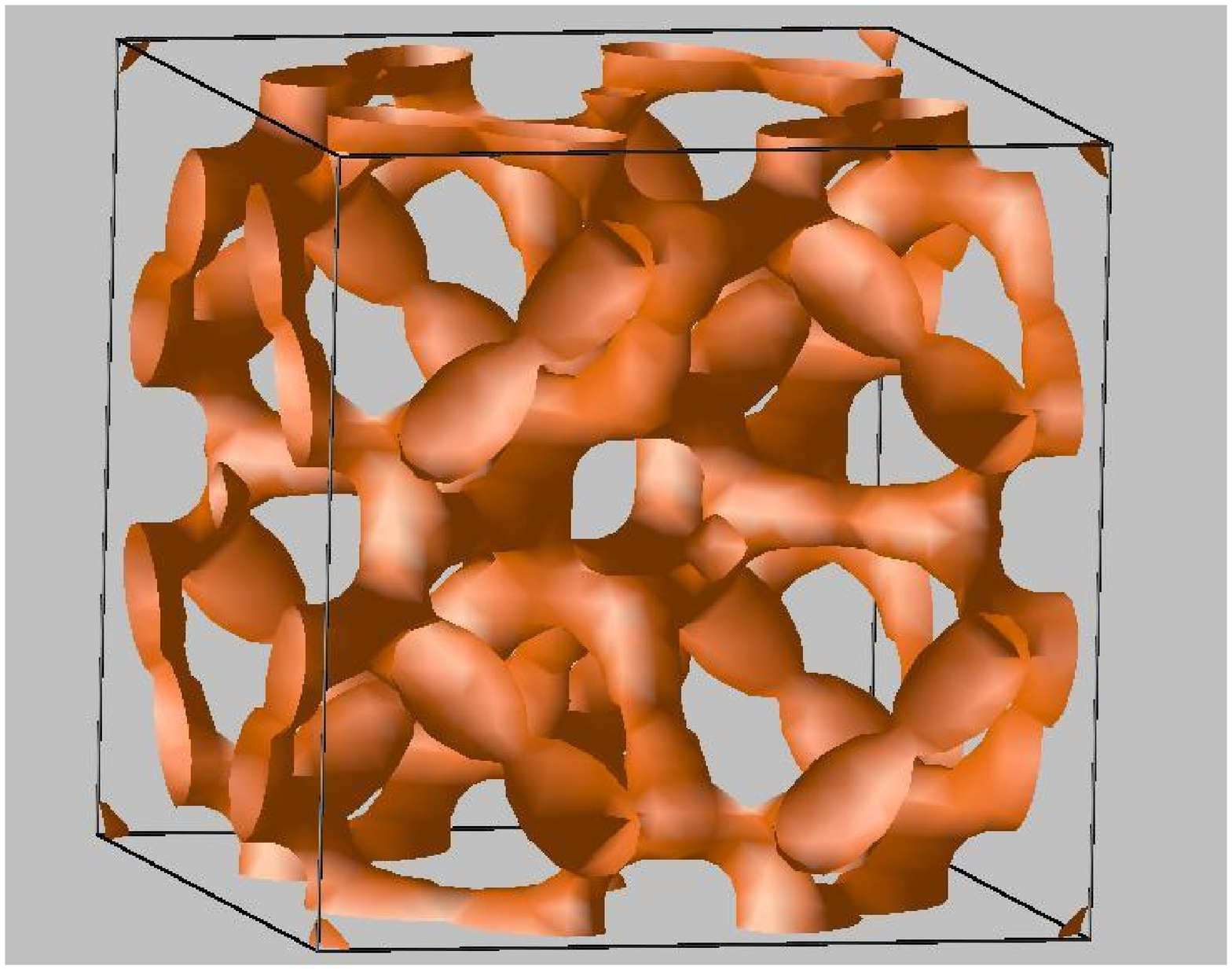,width=0.70\linewidth,angle=0,clip=}}
\vspace{0.2cm}
\caption{LDA Fermi surfaces of Y$_2$C$_3$.
}
\label{fs}
\end{figure}

The calculated electronic band structure and corresponding electronic
density of states (DOS) are given in Figs. \ref{bands} and \ref{dos},
respectively. The Fermi surfaces are displayed in Fig. \ref{fs}.
The band structure agrees qualitatively with a very recent report by
Shein and Ivanovskii, \cite{shein}
although there are noticable differences in detail, presumably related
to the different C-C bond length in that calculation.

\begin{figure}[tbp]
\centerline{\epsfig{file=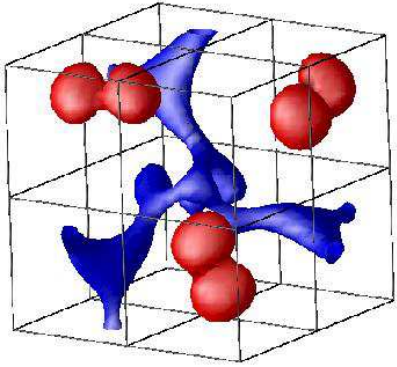,width=0.95\linewidth,angle=270,clip=}}
\vspace{0.3cm}
\caption{
This figure shows the charge density originating
from the valence bands near $E_F$ (see text)
in 1/4 of the primitive unit cell
around C (dumbells, red online)
and Y (interconnected network, blue online).
Isodensity surfaces correspond to $\rho ({\bf r})=$0.00112 e/$a_0^3$
for C and 0.00092 e/$a_0^3$ for Y. Small differences
between the charge density values were needed to improve visibility
of the two surfaces. For the same puprose, the parts of the
Y charge density surfaces, penetrating from the neighboring
unit cells, were removed. Note the metallic like Y derived density
and the C antibonding density.
}
\label{charge}
\end{figure}

The band structure shows a manifold of 12 C 2$s$ derived bands
extending from -14.9 eV to -6.2 eV, relative to the Fermi energy, $E_F$
(note that there are 12 C atoms in the primitive cell). This is followed
by a narrower manifold of 18 C 2$p$ bonding bands, associated with the
C dimers. This, in turn, is separated by a $\sim 2.5$ eV gap from a 
broad manifold of mixed C-C antibonding and Y character.
The Fermi level lies $\sim$ 1 eV from the bottom of this manifold.
Fig. \ref{charge} shows the contribution from states near $E_F$
to the charge density. The plot is for the energy range that
contributes 1 e per unit cell going down from $E_F$.
This electronic structure is consistent with a conventional picture
\cite{lee} of strongly covalently bonded C dimers embedded in a solid
that has metallic cohesion involving mixture of Y and antibonding C-C states.

As may be seen, there are several bands crossing the $E_F$. This leads
to a complex multisheet Fermi surface (Fig. \ref{fs}), with no
obvious strong nesting features. \cite{jack}
The density of states at $E_F$ is
$N(E_F)$ = 1.88 eV$^{-1}$ on a per formula unit both spins
basis, corresponding to a bare linear specific heat
coefficient, $\gamma_{bare}$ = 4.4 mJ/mol K$^2$. This is approximately
60\%
larger than the experimental value of $\gamma$ = 
2.8 mJ/mol K$^2$ (ref. \onlinecite{giorgi}),
but this comparison should be interpreted with caution since
the experimental stoichiometry was not well established.
We note that Th doped Y$_2$C$_3$ with $T_c$=17K (closer to the
18K reported for Y$_2$C$_3$ by Amano and co-workers), has a reported
experimental
$\gamma$ = 4.7 mJ/mol K$^2$, and that the DOS is peaked at the
stoichiometric composition, implying that off-stoichiometric samples
would likely have lower $\gamma$, both because of lower 
$\gamma_{bare}$ and presumably lower electron phonon enhancement.

\begin{figure}[tbp]
\centerline{\epsfig{file=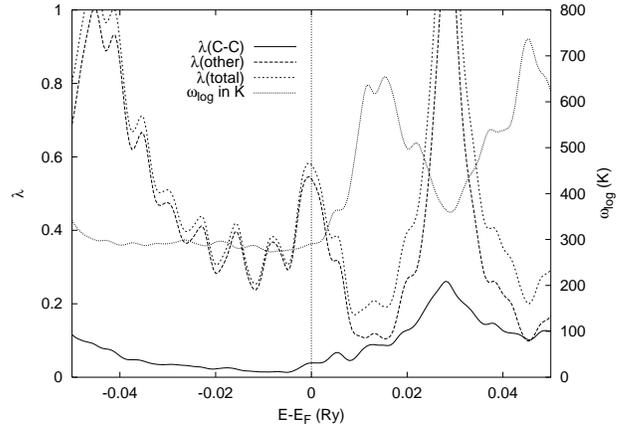,width=0.95\linewidth,angle=0,clip=}}
\vspace{0.2cm}
\caption{Mode $\lambda$ and average phonon frequency (see text) as a function
of energy. The heavy dashed (heavy solid) line denotes the full symmetry Y (C)
derived mode $\lambda$ as a function of energy, normalized as if these
modes are representative. The light dashed line is the total $\lambda$ with
this assumption. The light dotted line is the logarithmic average
of these two frequencies with $\lambda$, which would enter the
prefactor of the McMillan equation.
}
\label{lambda}
\end{figure}

Considering the apparent absence of strong nesting, and the
large unit cell (four formula units, 20 atoms), it is
expected that the zone center modes may be at least
semi-quantitatively representative for determining the
electron-phonon coupling. As a
further simplification, we focus on two modes
in order to obtain the gross features. These are the two
full symmetry modes related to the internal structural parameters.
As mentioned, these are
an Y dominated mode
at 175 cm$^{-1}$ and an almost pure C-C bond stretching mode at
1442 cm$^{-1}$. We determined the mode $\lambda$ for these two
modes using a frozen phonon method, in which LDA calculations were
done as a function of the phonon distortion, and the deformation
potentials calculated directly from the change in electronic structure
on the Fermi surface. The result is shown as a function of rigid
band Fermi energy in Fig. \ref{lambda}.
For this plot, it was assumed that the lower frequency Y mode
(which modulates both Y distances and Y-C distances) is representative
of all modes except the 6 C-C bond stretching modes, which were
assumed to be represented by the high frequency C-C mode.
Most likely, this is reasonably good for the bond stretching modes
but underestimates the phonon frequency
for the other modes, since the C dimers are lighter than an Y atom.
However, considering the crudeness of representing the system with two
full symmetry zone center modes, it hardly seems justified to attempt
corrections, and instead we view the result as semi-quantitative.

The calculated electron-phonon
coupling for the stoichiometric band filling is
$\lambda=0.6$ and is dominated by phonons other than the C-C bond
stretching; the C-C bond stretching provides less than 10\%
of the total $\lambda$. Inserting $\lambda=0.6$ into the McMillan
formula, with $\mu^*$=0.1 and $\omega$=175 cm$^{-1}$, we obtain
$T_c$=5K, which is fair agreement considering the crudeness of the
approximation employed.

What is significant is that $\lambda$ has a maximum at
the stoichiometric band filling. This is
related to the fact that the Fermi energy falls on a peak in the DOS.
The dependence of $\lambda$ on band filling and the difficult synthesis
of stoichiometric sesquicarbides provides an explanation for the
experimentally observed variability of $T_c$ depending on preparation
conditions.
Going to lower band filling, $\lambda$ decreases and then increases again
to a value exceeding $\lambda=1$, reflecting the
structure of the DOS, but this only happens quite far
down, corresponding to an electron deficiency of about 1 e per formula
unit. The C-C strecthing contribution is small throughout this region.
Going to higher electron count, $\lambda$ again decreases initially
and then increases, again to values exceeding $\lambda=1$. However,
in this case the C-C bond stretching contribution becomes significant
and as a result the logarithmic average phonon frequency also sharply
increases. In principle, this would mean that heavily electron doped
Y$_2$C$_3$ would become a high temperature superconductor, like MgB$_2$
and $A_3$C$_{60}$. However, this is based on a rigid band filling, which
is a highly unlikely scenario. Instead, population of C-C antibonding
states should be highly disfavored, and therefore it would be rather
difficult to chemically reach the region where this contribution to
$\lambda$ becomes large.

In summary, we report electronic structure and mode electron-phonon
calculations for Y$_2$C$_3$. The results are consistent with conventional
electron-phonon mediated superconductivity related to Y - C phonons.
The results show that the superconductivity
is qualitatively like in other metal carbide superconductors, and
not like MgB$_2$ or $A_3$C$_{60}$.

Work at the Naval Research Laboratory is
supported by the Office of the Naval Research.
The DoD-AE code was used for some calculations.

\end{multicols}

\begin{references}
\bibitem{jack}
The middle jack shaped section could have some nesting
between the flat parts along [111].
\end{references}
\end{document}